\documentclass[conference]{IEEEtran}
\IEEEoverridecommandlockouts
\usepackage{cite}
\usepackage{amsmath,amssymb,amsfonts}
\usepackage{algorithmic}
\usepackage{graphicx}
\usepackage{textcomp}
\usepackage{xcolor}
\usepackage{xspace}
\usepackage{url} 
\newcommand{\systemname}{TCU\xspace}
\newcommand{\longsystemname}{Trusted Compute Unit\xspace}
\usepackage{comment}
\usepackage{tikz}

\usepackage{cite}
\usepackage{amsmath,amssymb,amsfonts}
\usepackage{algorithmic}
\usepackage{graphicx}
\usepackage{textcomp}
\usepackage{xcolor}
\usepackage{xcolor}
\usepackage{cleveref}
\usepackage{comment}
\usepackage{booktabs}
\usepackage{atbegshi}

\newcommand{\hide}[1]{} %

\def\BibTeX{{\rm B\kern-.05em{\sc i\kern-.025em b}\kern-.08em
    T\kern-.1667em\lower.7ex\hbox{E}\kern-.125emX}}
\begin{document}

\newcommand\submittedtext{%
\footnotesize \textbf{Preprint}. This work has been accepted to the 2025 IEEE International Conference on Blockchain and Cryptocurrency (ICBC). © 2025 IEEE. Personal use of this material is permitted. Permission from IEEE must be obtained for all other uses, in any current or future media.}

\newcommand\submittednotice{%
\begin{tikzpicture}[remember picture,overlay]
\node[anchor=south,yshift=10pt] at (current page.south) {\fbox{\parbox{\dimexpr0.65\textwidth-\fboxsep-\fboxrule\relax}{\submittedtext}}};
\end{tikzpicture}%
}

\title{Trusted Compute Units: A Framework for Chained Verifiable Computations\\
}

\author{\IEEEauthorblockN{Fernando Castillo, Jonathan Heiss, Sebastian Werner, Stefan Tai}
\IEEEauthorblockA{\textit{Information Systems Engineering} \\
\textit{Technische Universität Berlin}\\
Berlin, Germany \\
\{fc,jh,sw,st\}@ise.tu-berlin.de}}

\maketitle
\submittednotice

\begin{abstract}
Blockchain and distributed ledger technologies (DLTs) facilitate decentralized computations across trust boundaries. However, ensuring complex computations with low gas fees and confidentiality remains challenging. Recent advances in Confidential Computing —leveraging hardware-based Trusted Execution Environments (TEEs)—and Proof-carrying Data—employing cryptographic Zero-Knowledge Virtual Machines (zkVMs)—hold promise for secure, privacy-preserving off-chain and layer-2 computations. On the other side, a homogeneous reliance on a single technology, such as TEEs or zkVMs, is impractical for decentralized environments with heterogeneous computational requirements.
 
This paper introduces the \longsystemname (\systemname), a unifying framework that enables composable and interoperable verifiable computations across heterogeneous technologies. Our approach allows decentralized applications (dApps) to flexibly offload complex computations to TCUs, obtaining proof of correctness. These proofs can be anchored on-chain for automated dApps interactions, while ensuring confidentiality of input data, and integrity of output data.
 
We demonstrate how TCUs can support a prominent blockchain use case, such as federated learning. By enabling secure, off-chain interactions without incurring on-chain confirmation delays or gas fees, TCUs significantly improve system performance and scalability. Experimental insights and performance evaluations confirm the feasibility and practicality of this unified approach, advancing the state of the art in verifiable off-chain services for the blockchain ecosystem.

\end{abstract}


\section{Introduction}
\label{sec:introduction}

In recent years, decentralized applications (dApps) and blockchain-based solutions have begun proposing off-chain computation mechanisms to address the high costs and privacy challenges of fully on-chain execution. For instance, a hospital might outsource AI-based diagnostic analytics to a specialized service that collaborates with multiple healthcare providers—similar to a federated learning (FL) setup. In this scenario, the hospital must trust that the external computation is performed correctly and confidentially, without exposing sensitive patient data on the blockchain or to unauthorized parties.

Such cross-organizational workflows often extend beyond a single “provider–consumer” pair, forming a directed acyclic graph (DAG) of dependent computations, thereby forming a cross-organizational workflow where one organization’s output can become another’s input. In these contexts, data integrity and the correctness of the underlying service computations are critical. Not only do organizations rely on accurate external data for sound decision-making, but they must also comply with stakeholder and regulatory demands to demonstrate the correctness of their claims. For example, the EU AI Act requires organizations to verify that their machine-learning applications conform to relevant regulations~\cite{laux2024three}.

However, achieving verifiability in computations across organizations is inherently challenging~\cite{liu2020privacy} due to the need to protect sensitive inputs while proving the correctness of computations. A naïve approach, such as re-executing computations with original inputs, introduces confidentiality risks, and verifying every computation on-chain translates to more gas fees and confirmation delays. Sensitive internal inputs, such as proprietary business data or personally identifiable information, cannot be exposed without violating confidentiality and security requirements~\cite{liu2016towards}. Moreover, prior research on verifiable computation\footnote{Some authors reserve 'verifiable computation' strictly for cryptographic proof systems. In this paper, we include TEEs under the broader category of trusted or verifiable techniques, while noting that TEEs rely on hardware trust assumptions.} has shown that no single technique or cryptographic construct can fully meet all desired properties of functionality and efficiency~\cite{yu2017survey,smart2023computing, bontekoe2023verifiable}. Instead, combining heterogeneous methods—such as zero-knowledge proofs and trusted hardware attestation~\cite{russinovich2024confidential}—remains a challenging yet promising direction, particularly in real-world scenarios where each organization maintains its own distinct technological stack.

Addressing these problems, we propose the \textit{Trusted Compute Unit (\systemname)} as a key component to enable interoperable and composable trusted computations across organizations. %
At its core, \systemname builds upon containers that wrap around services and execute data processing tasks within verifiable computation environments like zero-knowledge virtual machines (zkVM), or trusted execution environments (TEE),  to enable trusted chained computations. 
A Blockchain-based Program Registry is applied to enable decentralized and cross-organizational management and traceability of the \systemname's in- and outputs. 

In this paper, we make the following individual contributions:
\begin{enumerate}
    \item We present the \systemname as a unifying framework that enriches chained off-chain computations with verifiable computation and blockchain technology. \systemname{s} are modular and combinable components encapsulating service logic for verifiability and a foundational infrastructure. A blockchain-based registry enables cross-organizational traceability of workflow executions thereby also allowing for ex-post auditability. 
    \item We demonstrate the technical realization of \systemname using two fundamentally different technologies, i.e., TEE and zkVMs, and evaluate the impact achieving trustworthiness has on the system’s performance through initial experiments in a federated machine learning scenario, highlighting performance trade-offs between technology choice. 
\end{enumerate}

In the remainder of this paper, we introduce the Model in section 2 to introduce the system model, threat and requirements.  We present the \systemname Design in Section 3 and \systemname Technical Realization in Section 4. In Section 5, we describe the experiment-driven evaluation. We present related work in Section 6, and finally, conclude in Section 7.

\section{Model and Requirements}
\label{sec:model}
    
\begin{figure*}[t]
    \centering
    \includegraphics[width=0.8\textwidth]{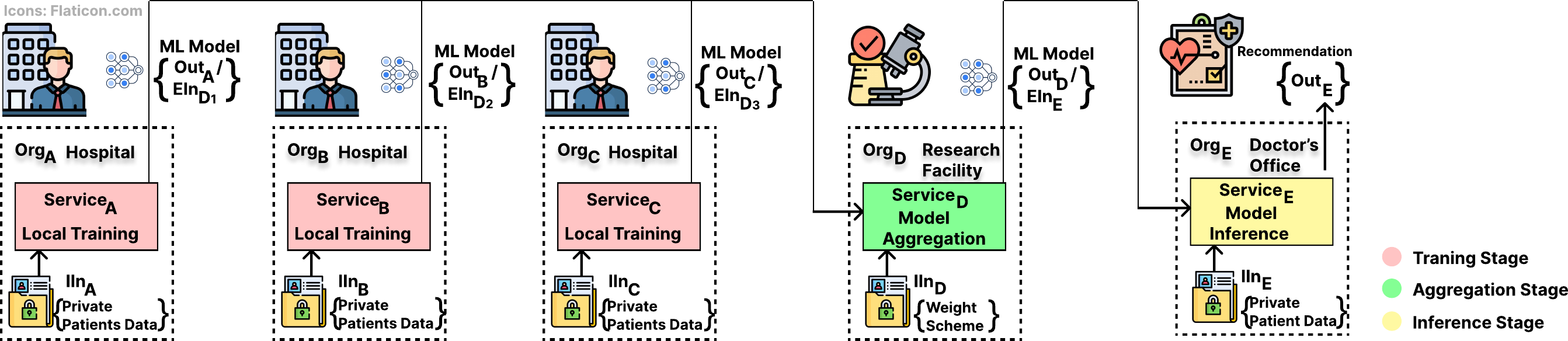}
    \caption{Illustration of a federated learning workflow. Organizations A, B, and C (Hospitals) perform local model training on private datasets ($IIn_{A,B,C}$), then send model updates ($Out_{A,B,C}$/$EIn_{D_{1-3}}$) to Organization D (Research Facility), which aggregates them  into a global model ($Out_D$/$EIn_E$) with a private weight scheme ($IIn_D$). Finally, Organization E (Doctor's Office) applies this aggregated model for making patient treatment recommendations ($Out_E$).}
    \label{fig:workflow_model}
\end{figure*}

In this section, we characterize  how cross-organizational service workflows operate in a \textit{decentralized} manner, and the threat model, define the core notions of \textit{computational} and \textit{workflow} integrity, discuss the confidentiality and verifiability challenges that arise, and derive the requirements our framework must fulfill.

\subsection{System Model: A Decentralized Service Workflow}
\label{sec:workflows}
We define cross-organizational service workflows as connected services, where each is executed by a different organization ($Org_i$, with $i$ the id for the location on the workflow). By \textit{decentralized}, we stress that no single coordinator or authority orchestrates the entire process. Instead, every organization controls and deploys its own service. Each service can consume:
\begin{itemize}
    \item \textbf{External inputs:} $EIn_i =\{EIn_{i}^1, EIn_{i}^2, \dots\}$, typically a \textit{set} of outputs from one or more predecessor services (e.g., $Out_i =\{Out_{(i-1)}^1, Out_{(i-1)}^2, \dots\}$, as different models to aggregate in a global model for FL).
    \item \textbf{Internal inputs} $IIn_i$, kept under the exclusive control of $Org_i$ and deemed confidential (e.g., proprietary weighting schemes for models or personally identifiable information).
\end{itemize}
Internal inputs are considered confidential, i.e., they must not be shared with external parties, while external inputs can be obtained by any organization.

In a typical workflow cycle, each \textit{required} service in the DAG is invoked at least once, so that its output can serve as external input ($EIn$) for any successors. Once a service $Out$ is not shared as $EIn$, we consider the workflow instance, conforming the particular DAG, complete. An example of such workflows can be seen in the context of decentralized AI application, as described in~\ref{fig:workflow_model}: %

Federated learning~\cite{zhang2021survey, heiss_advancing_2022,lee_VDFL_2024} comprises worker nodes and an aggregator node, each of which can be represented by a different organization. In each learning cycle, the worker nodes execute a machine learning task on local data representing confidential $IIn_i$ and collective learning results of the previous cycle representing $EIn_i$/$Out_{i-1}$. The resulting local models, representing $Out_{i}$, are sent to the aggregator node where they are combined into a global model which is returned to the worker nodes as $EIn_{i+1}$ for the next cycle. Additionally, the inference from a global model can be used to make recommendations for other patients.

Although a workflow may seem ``complete'' when terminal services produce their outputs, decentralized environments naturally allow new computations to \textit{extend} the DAG. For example in the FL scenario, a workflow execution represents a learning cycle, and a DAG composing all the computations could be recreated from the first local training until the last model aggregation after many learning cycles. Any output can serve as a fresh external input for a newly introduced service, underscoring the ongoing need for integrity guarantees across an evolving cross-organizational workflow.

\subsection{Threat Model}

We characterize the threat model by classifying each organization $Org_i$ as:
\begin{itemize}
    \item \textbf{Honest-but-negligent:} Executes its service  but may inadvertently misconfigure or misuse the code, generating incorrect outputs $(Out_i)$ or failing to implement the intended logic $P$ accurately.
    \item \textbf{Malicious:} Intentionally alters its logic $P$, data, or outputs to gain advantage (e.g., skipping computation in FL, not using the indicated ML model for inference). 
\end{itemize}

We assume no organization can forcibly read another’s internal inputs $({IIn}_{j\neq i})$; thus, data theft is not the primary threat. Instead, we focus on \textit{detecting} incorrect computations—whether caused by negligence or malice.

We do not consider side-channel or fault-injection attacks on hardware, nor do we assume any trusted setup. Other well-known security practices such as network security (e.g., TLS) to prevent eavesdropping in transit are assumed to be in place.

\subsection{Computational and Workflow Integrity}
Building on prior approaches for integrity and verifiability (e.g., \cite{bontekoe2023verifiable, heiss_2021_trustworthyPreprocessing}), we require that a program $P$ produce a proof $\pi$ demonstrating computational correctness under confidentiality constraints. Formally, given a \textit{proving key} $PK$, the execution is  $P(EIn, IIn, PK) \rightarrow (Out, \pi)$. Any verifier holding the corresponding \textit{verification key} $VK$ can then check the correctness via $verify(Out, \pi, VK)\rightarrow {\{0, 1\}}$, without needing access to the confidential inputs $IIn$.

\textbf{Computational Integrity (Single Service):} To define \textit{computational integrity}, we adopt the model presented in~\cite{heiss_2021_trustworthyPreprocessing} for trustworthy pre-processing off-chain data. A service program $P$ is executed on external input EIn and some internal input $IIn$ and returns output $Out$ such that $P(EIn, IIn) \rightarrow Out$. A malicious organization may benefit from corrupting either the program P or the inputs. In the former case, an organization creates a manipulated program P’ such that $P’ (EIn, IIn) \rightarrow Out’ \: | \: Out’ \neq Out$. Furthermore, an organization may manipulate the $EIn$ such that $P (Ein’, IIn) \rightarrow Out’ \: | \: Out’ \neq Out$ or manipulate $IIn$ such that $P (Ein, IIn’) \rightarrow Out’ \: | \: Out’ \neq Out$.

\textbf{Workflow Integrity (Chained Services):}
Because the output $Out_{i-1}$ from one service can become the external input $EIn_{i}$ of another, the \textit{entire} DAG preserves computational integrity, as long as $EIn_i$ is verified as part of $P_i$, step by step ensuring end-to-end \textit{workflow integrity}. Consequently, a final consumer can verify only the \textit{last} output without re-verifying all preceding services. In other words, they need \textit{not} check that \textit{every} intermediate output $(Out_{j<i})$ was produced by the intended program $(P_{j<i})$ on the correct inputs, even when $(IIn_{j<i})$ remain private—this guarantee is provided transitively through chained verifiable computations proofs, as each step verifies its input and computation, ensuring correctness propagates through the chain.

\subsection{Confidentiality and Verifiability Challenges}

Enforcing both \textit{computational} and \textit{workflow} integrity in a decentralized environment is non-trivial for several reasons:
\begin{itemize}
    \item \textbf{Preserving Internal Inputs:} Organizations cannot expose proprietary or personal data $(IIn_i)$ to external re-execution or naive on-chain logging or verification.
    \item \textbf{Preventing Undetected Code Changes:} If verification keys depended on hidden parameters or a trusted setup, an organization could surreptitiously modify the intended $P$, e.g., if the building process is not reproducible.%
    \item \textbf{Heterogeneous Execution Technologies:} Some organizations might use trusted execution (TEEs), while others might use cryptographic zero-knowledge systems. A unifying approach must handle both while preserving verifiability and low on-chain overhead.
\end{itemize}

\subsection{Requirements}

From the above setting and challenges, we derive the following key requirements for a cross-organizational trusted computation framework:

\begin{itemize}
    \item[\textbf{R1 -}] \textbf{Confidential Internal Inputs:} Each organization must retain confidentiality over its internal inputs $(IIn_i)$; no party should be forced to disclose or re-execute these sensitive data externally.
    
    \item[\textbf{R2 -}] \textbf{Verifiable Correctness of Outputs:} Any output $(Out_i)$ must be provably correct with  respect to the intended program $P$ on $(EIn_i,  IIn_i)$, without revealing private inputs. This includes  \textit{deterministic reference to the code of $P$ during verification}  so that any modification to $P$ invalidates existing proofs  and no hidden parameters can be introduced.

    \item[\textbf{R3 -}] \textbf{End-to-End Chainability:} Because the output of one service can become the input of  another, proofs of correctness must \textit{chain} through  the workflow. A final consumer should not need to re-verify  every step individually to ensure end-to-end correctness.

    \item[\textbf{R4 -}] \textbf{Cross-Organizational Traceability:} A transparent record (e.g., on a blockchain registry) must  indicate \textit{which service} produced \textit{which outputs},  under \textit{which code version}. This enables ex-post audits  and dispute resolution without exposing proprietary data.

    \item[\textbf{R5 -}] \textbf{Low On-Chain Overhead and Heterogeneity:} Minimizing on-chain interactions avoids high gas fees and prevents data leakage on public ledgers. Moreover, the framework must handle multiple verifiable computation technologies (e.g., TEEs or ZKPs) so each organization can choose the technology best suited to its needs without compromising overall verifiability.
\end{itemize}

In the following section, we present the \textit{\systemname} framework that addresses these requirements by combining off-chain verifiable computation technologies (TEEs or zkVMs) with a blockchain-based registry for deterministic code references and proof traceability.

\section{\longsystemname Design}
\label{sec:system_design}
To fulfill the requirements described in the previous section, we introduce our framework for \longsystemname{s} (\systemname{s}) by describing their key architectural components and the framework procedures.

\subsection{TCU Architecture}
\begin{figure}[h]
    \centering
    \includegraphics[width=0.95\columnwidth]{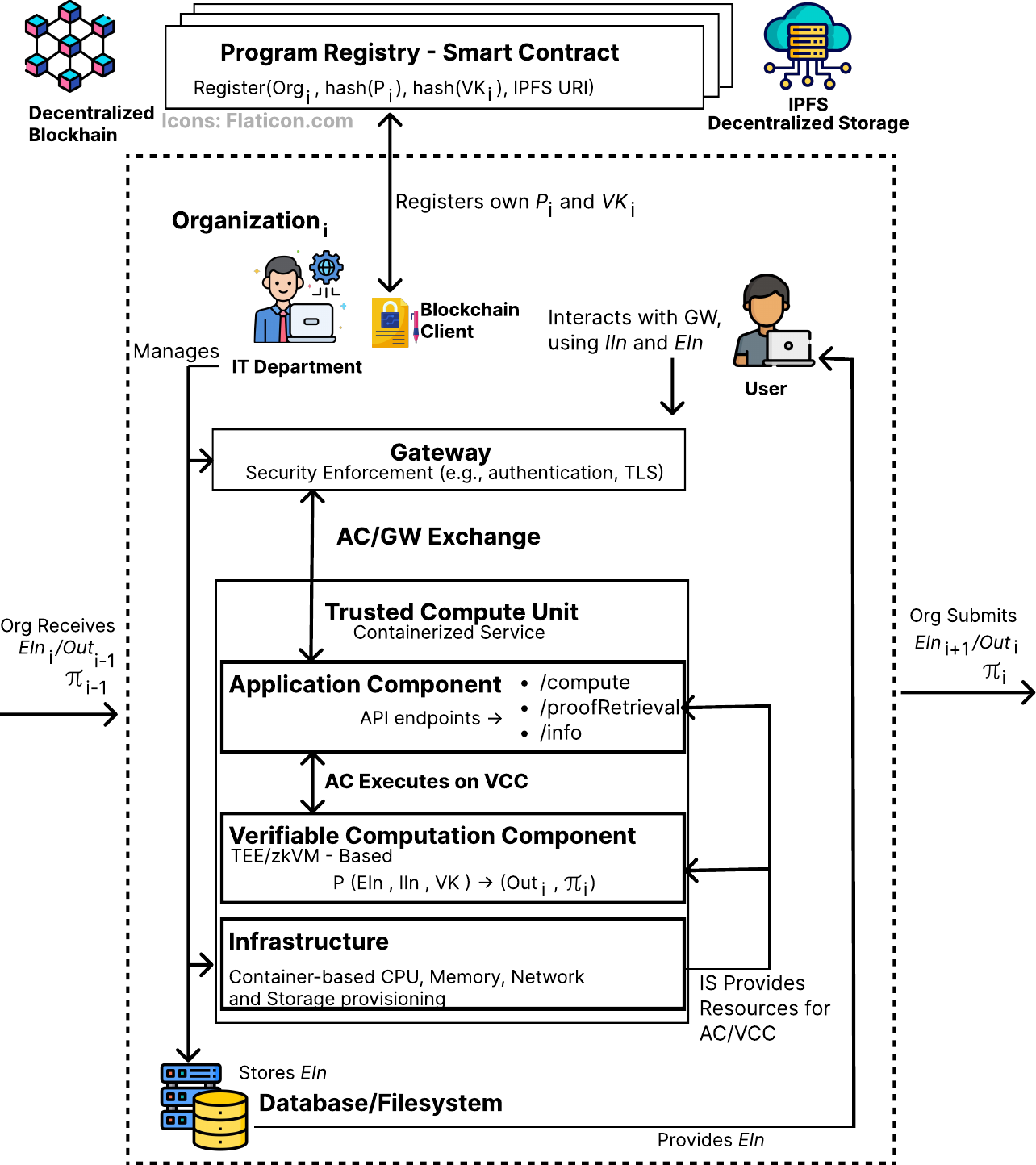}
    \caption{\systemname Framework Diagram
    }
    \label{fig:trustflow_system}
\end{figure}

As depicted in Figure~\ref{fig:trustflow_system}, the proposed framework advances cross-organizational workflows through two major architectural elements, the \systemname and the Program Registry (PR).

\systemname{s} are service units assumed to execute a program $P$ using a verifiable computation technology that allows a $TCU_i$ to create proofs $\pi$ with a \systemname-specific $PK$ and the successor $TCU_{i+1}$ to verify $\pi$ with the corresponding $VK$.

The \systemname can be modeled through 2 main components, the Application Component and Verifiable Computation Component, and a foundational Infrastructure for provisioning them.

\subsubsection{Application Component (AC)}
The AC provides a lightweight RPC interface for submitting computation tasks, retrieving results, and obtaining Program $P$ specifications. We assume that incoming requests have already passed through gateway-based authentication and authorization filters. By delegating these policies to external components, the TCU retains a simple boundary and focuses on core compute functionality. The minimal endpoints are a “\texttt{/compute}” call for scheduling tasks, and a “\texttt{/proofRetrieval}” route for obtaining proofs. A “\texttt{/info}” endpoint reveals metadata such as \systemname Program $P$ location for validation and used technology for verification. In practice, the \systemname only trusts requests that have already passed through the gateway’s authentication and authorization filters, keeping the component minimal and easily composable with other (micro)services. As illustrated in Figure~\ref{fig:trustflow_system}, its primary responsibility is to expose a minimal set of endpoints that interact with the \systemname{’s} Verifiable Computation Component.

\subsubsection{Verifiable Computation Component (VCC)}
The VCC can be modeled through three operations that are all executed (via the AC “\texttt{/compute}” call) within the same verifiable computation technology and a key pair consisting of the proving key (PK) used to create proofs $\pi$ by the $TCU_i$ and the verification key (VK) used to verify $\pi$ by the successor $TCU_{i+1}$. 

The (1) \textit{External Input Verification} (EIV) takes the proofs from its predecessor TCUs as input and verifies them using the respective verification keys. With that, the computational integrity of the previous \systemname operations can be confirmed. 
On successful EIV, the (2) \textit{service is executed}. It takes the $IIn$ and the verified $EIn$ and returns the computational output.
Finally, the \systemname (3) \textit{commits to the internal inputs} which is necessary to bind the inputs to the service execution. 
The \systemname execution returns the computational output to the AC (so it's retrieved in the “\texttt{/proofRetrieval}” endpoint), the commitment to the internal inputs, and the proof of computational correctness. The $VK$ of a \systemname is used for EIV by the successor \systemname.

\subsubsection{Infrastructure (IS)}
The IS includes provisioning CPU, memory, and storage resources, orchestrating container deployments, and managing lifecycle events such as rolling updates. Container orchestration frameworks (e.g., Kubernetes) or virtualization managers typically provide these capabilities.

The IS supports features like automatic scaling, which starts additional TCU instances in response to rising workloads, or rolling upgrades to deploy patches with minimal downtime. It also enforces isolation boundaries between routine and verifiable computations, ensuring the \systemname remains shielded from potential interference. Deterministic builds and cryptographic checksums can verify that the deployed code matches the audited binaries (matching the “\texttt{/info}” endpoint), mitigating supply-chain attacks~\cite{lamb2021reproducible} and preserving trust over time.

In practice, the Infrastructure is typically administered by the organization’s internal IT Department or DevOps teams, whether running on-premises or via a cloud provider. While the Figure~\ref{fig:trustflow_system} shows this infrastructure ‘inside’ the organization, its physical location can vary according to each team’s operational policies without affecting the \systemname's verifiability or confidentiality guarantees.

\subsubsection{Program Registry}
 We introduce a smart contract-based program registry (PR), on a decentralized blockchain, that stores the specifications of the program $P$. By anchoring program $P$’s logic, configuration, and cryptographic references on an immutable ledger, external verifiers can confirm exactly which code and parameters generated any given proof—even after the TCU has gone offline or been replaced. This approach decouples ephemeral TCU computations from the enduring record of the application-specific program, allowing auditors or external services to validate that a proof indeed corresponds to the original, trusted version of $P$ (matching the “\texttt{/info}” endpoint). By maintaining a tamper-proof ledger entry, we ensure that verifiability persists long-term, preserving integrity and trust in the computed results.

\subsection{\systemname Life Cycle}
As depicted in Figure~\ref{fig:trustflow_system}, the service lifecycle consists of a one-time setup and recurring “\texttt{/compute}” calls.
As a prerequisite, we assume that the Program Registry is deployed to a decentralized public infrastructure as provided by smart contract-enabled blockchains.

\textbf{Setup:}
During the setup, the \systemname is instantiated and the program $P$ registered in the registry. The \textit{instantiation} involves the specification of the logic as $P$, the compilation into an executable and provable format. This ensures that the AC has a well-defined program $P$ to execute within the VCC. %
Once integrated, the \systemname-specific proving and verification keys are generated for the VCC. 

For \textit{registration}, the organization registers $P$ into the Program Registry, the \systemname{’s} identifier, and the $VK$.

\textbf{Operation:} Once the setup is complete, the TCU can operate in a service workflow. On receiving the  “\texttt{/compute}” call, with $EIn_{i-1}$ from a predecessor \systemname, the AC checks the validity of the predecessor $TCU_{i-1}$, from the corresponding record on the PR. 
Then, the AC calls the~\systemname$_i$'s VCC to execute the program $P$ with the operations on the internal and external inputs, with the successful EIV as a prerequisite for the service execution and the conclusive internal input commitment. 
The execution returns a proof of computational correctness $\pi$ with the output and the internal input commitment $C(IIn)$. Then an authorized User submits $\pi$ and $C(IIn)$ to the successor Organization's \systemname, using the “\texttt{/proofRetrieval}” endpoint through the AC.%
\\

In our framework, we use the term “\longsystemname{}” (\systemname), reflecting how trust emerges from satisfying the framework’s requirements within a self-contained, manageable unit-like service interface. Besides the $VK$, each \systemname offers a deterministically generated reference to $P$ for verification (\textbf{R2: Verifiable Correctness of Outputs}) and anchors that reference and the $VK$ on a Program Registry (\textbf{R4: Cross-Organizational Traceability}), ensuring any code changes become self-revealing. The AC presents a uniform API so heterogeneous verifiable computation technologies—TEEs or zkVMs—can be integrated without modifying service endpoints (\textbf{R5: Low On-Chain Overhead and Heterogeneity}). Meanwhile, the VCC preserves confidential inputs (\textbf{R1: Confidential Internal Inputs}) and chains correctness proofs across computations (\textbf{R3: End-to-End Chainability}). Lastly, because only sink computations need on-chain verification (if required by a dApp), the on-chain overhead remains minimal (\textbf{R5} again), while the IS orchestrates off-chain deployments seamlessly. This design unifies code provenance, proof generation, and cross-technology composability in a portable, tamper-evident framework for verifiable off-chain services.

\section{Technical Realizations}
\label{sec:tech_realization}
In this section, we describe how \systemname{s} can be realized with current Container Orchestration Frameworks for the IS, with Trusted Execution Environments (TEEs) and Zero-knowledge Virtual Machines (zkVMs) for the VCC, and lightweight web frameworks for the AC. 
While we consider these technologies as candidates, we would like to underline the generality of \systemname which may be used as a blueprint for further technologies, in particular for other Verifiable Computation Technologies, e.g., FHE and SMPC~\cite{yu2017survey,smart2023computing, bontekoe2023verifiable}. 
Finally, we outline how the PR can be realized with blockchain smart contracts. 
\subsection{Infrastructure with Orchestration Framework}

In practice, this layer leverages container orchestration frameworks (e.g., Kubernetes or Argo Workflows) to provision and lifecycle-manage \systemname instances. For TEE-based \systemname{s} (e.g., Intel TDX or AWS Nitro Enclaves), the corresponding container images include both the application logic and enclave drivers, deployed only on nodes that support enclave-capable hardware. For zkVM-based \systemname{s} (e.g., Risc0), the container images embed the zkVM runtime and deterministic ELF binaries. The IS ensures correct scheduling (e.g., via node labels), secure distribution of proving and attestation keys, and rolling updates to patch or replace \systemname containers with minimal disruption. In the case of TEEs, it also facilitates remote attestation workflows, where the container attests to a trusted authority (e.g., Intel TA or AWS Nitro) prior to accepting inputs. For zkVMs, container-level integrity checks (e.g., Docker Image signing) ensure the correct zkVM runtime is loaded. Finally, the IS integrates with DevOps pipelines (e.g., reproducible builds, image signing) to maintain a consistent, verified environment for \systemname operations.

\subsection{Verification Computation Component with Trusted Execution Environments}
Trusted Execution Environments (TEEs) are secure hardware components that safeguard data and code from external tampering and disclosure\cite{munoz2023survey}. 
Programs executed within TEEs operate within isolated and/or encrypted memory regions, shielding the content even from the hardware owner and ensuring the integrity of computations conducted within. 
While TEEs also enable confidential computation, i.e., protecting \textit{data in use} from the executor, we leverage its ability to make internally executed programs externally verifiable.

\textit{Remote Attestation} enables external parties to verify the integrity of a Trusted Execution Environment (TEE)'s internal state and the authenticity of messages from within it. TEE-enabled machines have a machine identity key embedded into the hardware during manufacturing. Using this key, each TEE instance generates an identity certificate that can be externally verified through a Public Key Infrastructure (PKI). These keys are used to sign measurements that represent a complete snapshot of the TEE’s internal state at boot. When a remote attestation request is made, the TEE returns signed measurements that can be reconstructed outside the TEE to verify the integrity of its internal state. These measurements can include a reference to a specific container, whose correctness can be validated with a deterministic build system~\cite{lamb2021reproducible}, thereby enabling verification of a trustworthy computation of the program $P$.
This general model of TEEs represents the essential concepts of confidential VMs, or containers\cite{confidential2022common}, like AWS Nitro~\cite{brossard2023private} which we use for the \systemname realization.

During the \textbf{setup}, the \systemname's program logic with the verification keys of the predecessor \systemname{s} are specified and compiled during the initialization of the TEE. On initialization, the measurements of the internal state, i.e., the \systemname's binaries, are created and signed through the TEE-specific key for remote attestation. This attestation report and the TEE's public key and specific container are registered at the Program Registry. The attestation represents a commitment to the \systemname allowing other workflow parties to check the integrity of the TEE's internal state whereas the TEE-specific public pair represents the verification key. %
    
In the case of the workflow \textbf{operation}, the $EIn$ and $IIn$ are provided through the AC, the host of the TEE, where the \systemname logic is executed. After executing the three \systemname VCC operations, the computational output $Out$ and the commitment to the internal inputs $com(IIn)$ are signed with the TEE's private key. The signature representing the proof of computational integrity can be verified using the corresponding TEE's public key.%

\subsection{Verification Computation Component with Zero-Knowledge Virtual Machines}
Zero-Knowledge Virtual Machines (zkVMs) leverage non-interactive zero-knowledge proof (ZKP) protocols to make the computational integrity of programs executed inside the zkVMs independently verifiable. 
ZKPs can encode program logic in mathematical constraint systems called circuits. 
Computational integrity can be asserted if a valid, input-specific variable assignment of the constraint system can be found. 
To enable external verifiability without disclosing computational inputs, elliptic curve cryptography applies where proofs are constructed and verified with a circuit-specific key pair. %
ZkSTARKs, as described in~\cite{ben2019scalable}, define a class of such protocols, which is typically used in zkVMs, characterized by fast proof generation and a transparent, deterministic setup free of trust assumptions, enabling verification of $P$’s computation.

Different from application-specific circuits, zkVMs encode the instruction set of virtual machines in circuits.
This allows for executing programs in a format compliant with the zkVM's instruction set. 
Risc0 \cite{bruestle2023risc}, for example, is a zkVM built upon the RiscV instructions and, hence, allows executing programs deterministically compiled to general Executable and Linking Format (ELF) binaries. 
The execution of an ELF binary file returns a cryptographic proof that can be verified in any other Risc0 zkVM through a reference to the binary, called the ImageID. 
For the following, we assume Risc0-enabled \systemname{s}. 

During the \textbf{setup}, the \systemname logic is specified in a high-level language and compiled into an ELF binary file. From that, the ImageID can be created which is a cryptographic (hash-based) representation of the initial zkVM memory state produced when the ELF binaries are loaded. The ImageID allows the zkVM of the successor organization to verify that the computational proof has been generated by the expected ELF binary. 
    The ImageID corresponding to the verification key and the ELF binary representing the \systemname commitment are made available on the Program Registry. 
    
In the case of \textbf{operation}, the AC provides the $IIn$, $EIn$, and ELF binary to the zkVM's executor which runs the ELF binary and records the session as complete snapshots of the state of the zkVM throughout the execution. 
    Based on that, the Receipt is created which serves as proof of computational integrity.
    The Receipt contains the computational outputs, the execution's imageID, and the seal, a cryptographic artifact that attests to the validity of the outputs and imageID. 
    The Receipt can be verified with the original ImageID in the VCC of the successor organization.     
\subsection{Application Component with Containers}

For a real-world deployment, the AC is realized as a lightweight containerized service exposing RPC endpoints (e.g., \texttt{/compute} and \texttt{/proofRetrieval}). An external gateway (such as Envoy or Istio) is delegated with the handling of advanced features like authentication tokens, TLS termination, and rate limiting before requests reach the \systemname.

Inside the \systemname container or enclave, a minimal web framework (e.g., a Flask service) receives validated requests and forwards them to the VCC. When the \texttt{/compute} endpoint is called, the AC packages $EIn$ along with any local $IIn$ and invokes the verifiable code inside the TEE or zkVM runtime. Once the computation finishes, on the VCC, it returns the resulting output  with a verifiable proof.

For chained trusted computations, the AC also records a proof reference or \systemname identifier on the Program Registry (also available at the \texttt{/info} endpoint), enabling cross-organization independent verification. Similarly, the \texttt{/proofRetrieval} endpoint serves as a lightweight retrieval mechanism for downstream consumers to obtain existing proofs on demand. Because business logic and advanced security checks reside outside the \systemname, this component remains easy to adapt and integrate into existing DevOps pipelines or microservice meshes, ensuring each \systemname instance can be deployed with minimal friction.

\subsection{Program Registry with Blockchains}
Blockchains are decentralized systems designed to resolve trust issues among collaborating parties without depending on trusted third parties (TTPs). %
Submitted transactions are redundantly processed through a collectively executed consensus protocol, and agreed-upon transactions are securely recorded in an immutable, append-only transaction history.

We use smart contracts to technically realize the Program Registry, e.g., those based on Ethereum Virtual Machine~\cite{wood2014ethereum} or  WebAssembly~\cite{tara2019evolution}. %
Each organization is assumed to have its own blockchain account represented through a public-private key pair. 
As transactions are by default authenticated with the organization's account keys, transactions can be associated with the organization's blockchain account.
While this hides the organization's identity behind the account representation, it allows to associate \systemname{s} if the mapping of the account keys and the organization's identity are known.  

As blockchains suffer from storage limitations, we propose keeping large artifacts off-chain, in particular $P$, in distributed file storage like IPFS~\cite{IPFS_Benet_2014} and only storing a hash-based reference on the blockchain to preserve the artifacts' integrity and transparent validation. 
Such off-chain storage patterns are well-known and, for example, described in~\cite{eberhardt2018off}.

\section{Evaluation}
\label{sec:Evaluation}
\label{sec:Discussion}
In this section, we evaluate \systemname by assessing trustworthines, technical and performance concerns in the context of a FL scenario, and discuss open security-related issues. %

\subsection{Experiment-driven Evaluation}
To technically evaluate the \systemname framework, we implement the \systemname's VCC using Risc0\footnote{https://dev.risczero.com/api/zkvm} as a zkVM, and AWS Nitro Enclave\footnote{https://aws.amazon.com/de/ec2/nitro/} as a VM-based TEE, and the PR in EVM-based~\cite{wood2014ethereum} for smart contracts execution. The AC is a light web server API on Flask\footnote{https://flask.palletsprojects.com/en/stable/} containerized on a Dockerfile\footnote{https://docs.docker.com/reference/dockerfile/}, and the IS is managed with AWS Nitro Hypervisor. The experiments were executed in an AWS EC2 instance of type c5.4xlarge, with 16 vCPUs and 32 GB of memory. %

We evaluate the \systemname in a federated learning (FL) scenario as introduced in Section~\ref{sec:model}. 
In FL, the \systemname helps to mitigate \textit{trustworthiness concerns} derived from aggregation attacks and model poisoning~\cite{xia2023poisoning} through intervened proofs of computational integrity~\cite{heiss_advancing_2022, lee_VDFL_2024}.

For the local learning node, we implement a simple neural network with two hidden layers, using stochastic gradient descent as the optimization method, as the $P$ logic of a \systemname, while doing dataset authentication with a signature from the node and verifying that the base global model to use comes from a \systemname aggregator. In the case of aggregation, we implement federated averaging~\cite{nilsson2018performance} inside the \systemname which takes the average of local model updates of the worker nodes and additionally verifies that the local models come from a \systemname. 
Without compromising the generality of our approach, the FL scenario allows us to address the previous concerns about trustworthiness, while evaluating performance in the following experiments E1-E4\footnote{Repository for experiments: https://github.com/ferjcast/TCU}: %

\begin{itemize}
    \item [E1] We set up and deploy \systemname{s} with different VCCs, TEE and zkVM-based, for a worker and aggregator node and measure execution times and transaction costs. %
    \item [E2] We run the same compute job, i.e., learning and aggregation, in a TEE and a zkVM-based VCC and alternate \systemname$_{i-1}$ to verify proofs from TEEs or zkVMs.%
    \item [E3] We increase the dataset volume, in the range of 20 to 10240 rows, to measure how it affects the authentication of the dataset and learning phase. %
    \item [E4] We increase the number of learners, in the range of 2 to 800, to measure how the proof verification time grow inside of the aggregator VCC.%
\end{itemize}

For E1 to E4 we switch between a zkVM-based and TEE-based aggregator and learners respectively.

\textbf{Results and Evaluation:} The practical implementation of the \systemname, which addresses the \textit{trustworthiness concerns}. By chaining proofs of computational integrity together, such guarantees extend across multiple heterogeneous service executions along the workflow (\textbf{R3, R4, R5}), $IIn$ can not be tampered unnoticeably either as Data for the Local Model or the Weighting Scheme, nor $EIn$ as a Local or Global Model. At the same time the \systemname information is maintained on the blockchain-based PR alongside the storage of $EIn$ on each organization's side (\textbf{R2}). 

The \systemname protects against input tampering attacks, while guaranteeing input confidentiality, (\textbf{R1, R2}) inherently by using verifiable computation and against formal incorrectness by storing commitments to \systemname's $P$ in the PR, so the $P$ reference to $VK$ can be created on the verifier side.

Cheating organizations can be held accountable if irregularities in shared data are detected, e.g., through plausibility checks. 
The integrity of $IIn$, $EIn$, and \systemname program $P$ can then be verified by inspecting $P$ using the PR. 
This helps in dispute cases or audits to solve conflicts ex-post.

\begin{table}[h]
\centering
\caption{Container Image Size for Each Scenario.}%
\label{tab:vcc-artifacts}
\resizebox{0.95\columnwidth}{!}{%
\begin{tabular}{l|l|l|l}

\hline
\textbf{Node Task} & \textbf{$VCC_i$(Computing)} & \textbf{$EIn_{i-1}$ Verified} & \textbf{Image Size} \\
\hline
Learner & zkVM & TEE & $\sim$6.4 GB\\ %
Learner & zkVM & zkVM & $\sim$6.4 GB\\ %
Learner & TEE & TEE & $\sim$1.8 GB\\ %
Learner & TEE & zkVM & $\sim$6.4 GB\\  %
Aggregator & zkVM & TEE & $\sim$6.4 GB\\ %
Aggregator & zkVM & zkVM & $\sim$6.4 GB\\ %
Aggregator & TEE & TEE & $\sim$1.8 GB\\ %
Aggregator & TEE & zkVM & $\sim$6.4 GB\\  %
\hline
\end{tabular}
}
\end{table}
Across all scenarios, on-chain PR deployment consumes about 979k gas, and registering a TCU costs roughly 265k gas, regardless of dataset or model sizes. Off-chain TEE attestation remains negligible at \(\sim 0.01\,\mathrm{s}\). However, Table~I indicates container image size is inflated by the requirement of the Risc0 framework library to be included in the image (for proving or verifying with the zkVM), reflecting the difference between bundling the specific binaries for a computation  versus adding the specific cryptographic runtime.

\begin{figure}[htb]
    \centering
    \includegraphics[width=0.76\columnwidth]{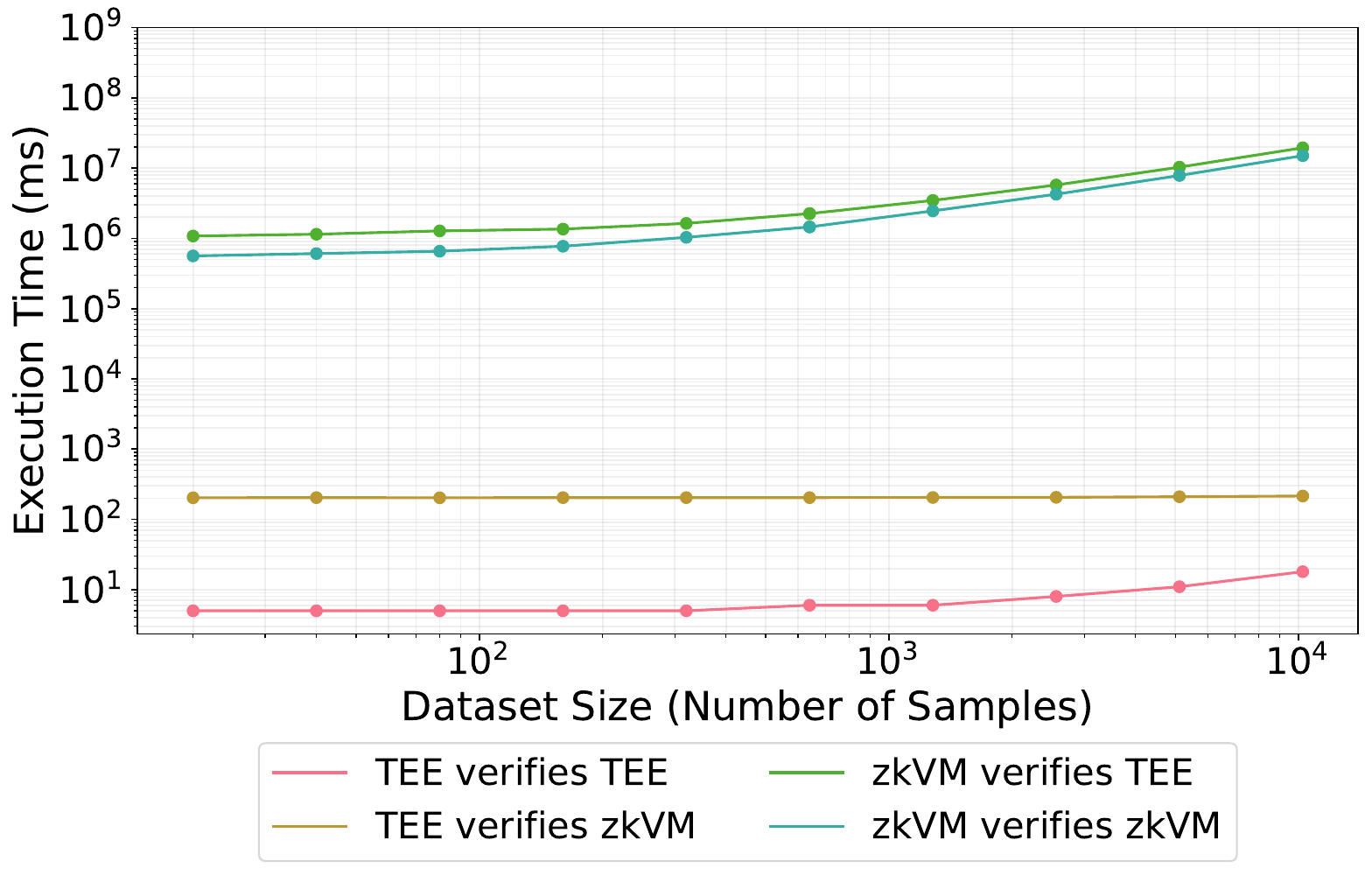}
    \caption{Impact of data volume ($IIn_i$ size) in Local Training scenarios. VCC$_i$-type verifies EIn$_{i-1}$-type.}
    \label{fig:training_nodes}
    \centering
    \includegraphics[width=0.76\columnwidth]{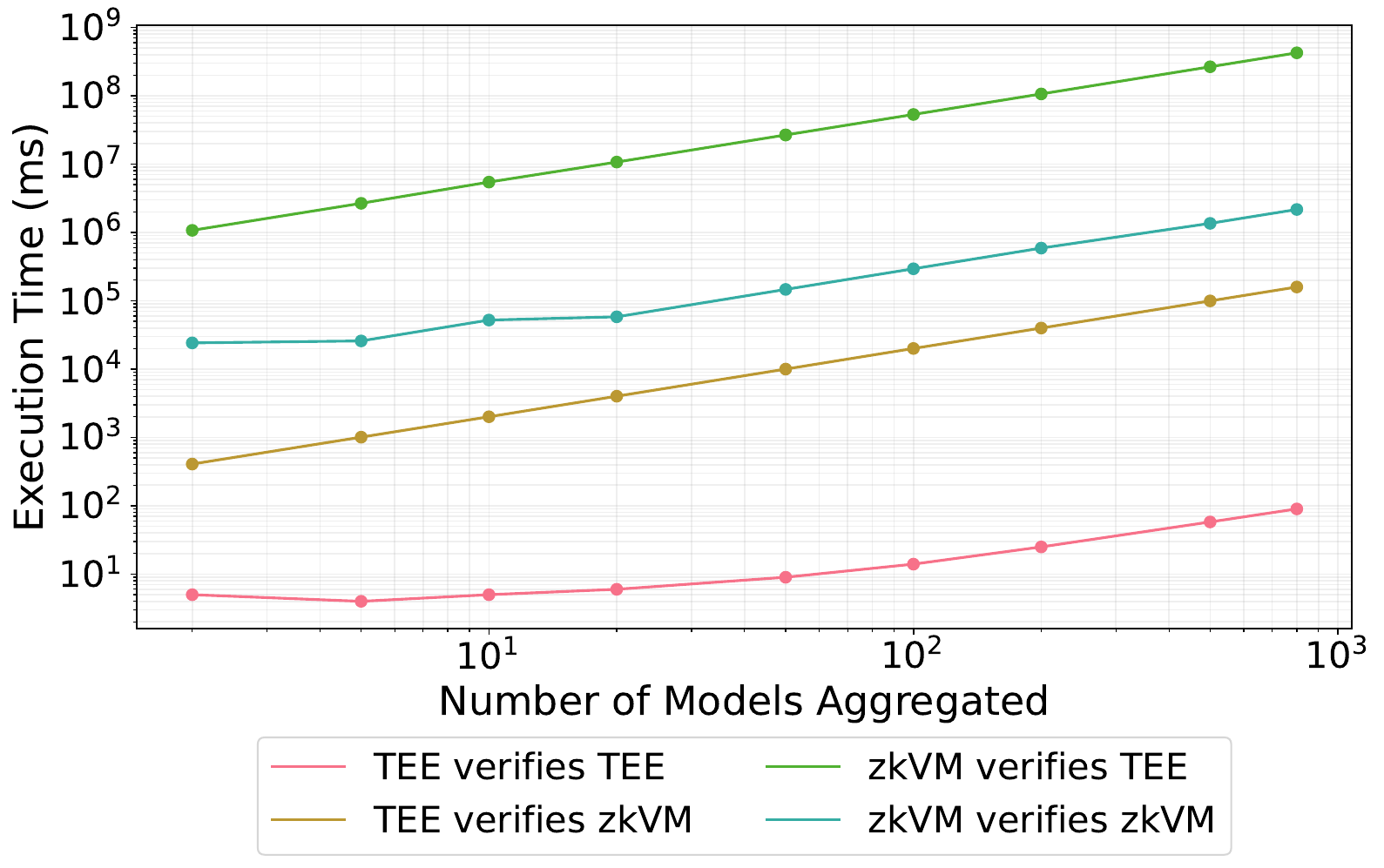}
    \caption{Impact of total number of models ($EIn_i$ set size) in Aggregation scenarios. VCC$_i$-type verifies EIn$_{i-1}$-type.}
    \label{fig:aggregation_node}
\end{figure}

For local training on \(n\) data samples, ~\Cref{fig:training_nodes} shows TEE verification can exceed the training cost for smaller \(n\) (up to \(\sim 2400\)), while zkVM overhead escalates faster beyond \(n=320\). Model aggregation scales linearly with the number of local models (~\Cref{fig:aggregation_node}), each requiring a single verification plus an averaging step. For both types of nodes, TEEs run the same compiled instructions as regular binaries, whereas zero-knowledge systems must encode logic into cryptographic operations—leading to \textit{multiple orders of magnitude} difference of execution time.

\subsection{Discussion}
Complementing the previous performance evaluation, we now revisit key aspects of \systemname's for its real world appliaction. %

A significant finding for both scenarios, learning and aggregation, is that the verification of a \systemname from a previous \systemname is more efficient when both the learner \systemname and the aggregation \systemname share the same technology type. %
These results can help organizations manage resource allocation more effectively as workflows expand. For example, load distribution strategies can be optimized throughout the workflow, given that TEEs generally exhibit better performance compared to zkVMs. %
Understanding these factors and developing tailored strategies to address specific use case constraints will significantly enhance the practical implementation of \systemname. 
In practice, whether a \systemname-based service can tolerate higher proof-generation overhead or deferred execution depends on its operational requirements. For instance, model training may only occur weekly or monthly, making it acceptable to use a more computation-intensive verifiable environment if the update cadence is low. By contrast, inference services demand quicker turnaround. This flexibility allows each organization to pick the most suitable verifiable technology in line with the service’s real-time or batch-processing needs, without compromising the overall trust guarantees.
This initial experimentation can guide organizations in efficiently deploying \systemname{s} across heterogeneous VCCs.

\section{Related Work}
\label{sec:relatedWork}

In this section, we review the related work to \systemname{s}. We include data provenance models, blockchain-based approaches, and existing cooperative services.%

Models for provenance~\cite{moreau2011open} and the PROV\footnote{https://www.w3.org/TR/prov-overview/} specification from W3C have been designed with only a focus on data, not on the computations and processes producing the data.
Hence, systems in support of data provenance~\cite{gehani2012spade,baracaldo2017securing,malik2018productchain} exist, but all only consider the data and not the computations. %
Some approaches leverage the blockchains to enhance the trust in data provenance~\cite{mertens2024smart}, or also provide service composition information as part of the data provenance record~\cite{ardagna2020blockchain,corradini2021chorchain}. Additionally, some recent efforts like ZkTLS~\cite{kalka2024comprehensive} focus on applying zero-knowledge to the TLS session, allowing clients to verify certain properties of the session without revealing sensitive details.
However, these approaches do not have verifiability for the computations generating the data.
Hence, approaches like the \systemname would add and enhance data provenance.

Similar approaches for verifiable workflows already exist, for instance, some using only TEEs~\cite{delignat2023should,howard2023confidential}.
Authors in ~\cite{toldi2023blockchain}, also use commitment in the blockchain of a hashed proof, represented as $EIn$ in the \systemname, and leveraging ZKP-based computational correctness guarantees, but their approach has higher gas costs and only uses zkSNARKS.
Hence, existing approaches do not consider the adaptability to other types of VCCs, and thus, do not enable organizations to choose between the trade-offs of each verifiable computing technology, unlike \systemname{s}.
Other noteworthy ZKP-based workflow systems do not focus on the confidentiality aspect, for example, using blockchains as the verification layer~\cite{ramezan2024zk} or store public proofs on-chain, thus introducing loss of confidentiality~\cite{heiss2023verifiable,heiss_2021_trustworthyPreprocessing}.

Lastly, some approaches also tackle the reproducibility problem without na\"ive re-execution using secure network provenance, both without and with the requirement of trusted hardware components~\cite{zhou2011secure,zhou2022audiwflow,taha2015trusted}. %

\hide{
\begin{table}[htbp]
\centering
\caption{Comparison of \systemname with Other Approaches}
\label{tab:comparison}
\begin{tabular}{lcccc}
\hline
Feature & \systemname & Blockchain-only & TEE-only & ZKP-only \\
 & (Proposed) & Approaches & Approaches & Approaches \\
\hline
Computational Integrity & $\checkmark$ & Partial & $\checkmark$ & $\checkmark$ \\
Verifiability & $\checkmark$ & Partial & $\checkmark$ & $\checkmark$ \\
Confidentiality & $\checkmark$ & Partial & $\checkmark$ & $\checkmark$ \\
Proof Chaining & $\checkmark$ & $\times$ & Partial & Partial \\
Workflow Traceability & $\checkmark$ & $\checkmark$ & Partial & Partial \\
Technology Agnostic & $\checkmark$ & $\checkmark$ & $\times$ & $\times$ \\
\hline
\end{tabular}
\end{table}
}
While these approaches offer valuable insights into various aspects of secure and verifiable cooperative services, they often focus on only one of those specific elements such as confidentiality, verifiability, or traceability. %
Unlike previous approaches, that are technology dependent, \systemname is a flexible framework that can operate technology agnostic with its abstraction of the \systemname with the possibility of chaining verifiable computations.%

\section{Conclusion}
\label{sec:Conclusion}
We introduced the \systemname, a framework for a verifiable service workflow system for trustworthy cross-organizational data sharing. 
Using \systemname{s} allows organizations to verify and demonstrate the computational correctness of a service workflow. The containerized design of \systemname{s} integrates seamlessly with existing services and orchestration pipelines, facilitating ephemeral deployments that are reproducible and auditable.

Our experiments confirm the \systemname's interoperability and composability across TEEs and zkVMs, offering flexible trust solutions for cross-organizational workflows. Thus, the \systemname emerges as a pioneering approach to next-generation verifiable off-chain computations. Future research will broaden \systemname's scope to incorporate additional TEE variants (e.g., Intel TDX) and other zkVMs (e.g., Nexus), as well as alternative verifiable computation paradigms such as Fully Homomorphic Encryption and Secure Multi-Party Computation, further reinforcing \systemname's vision of an agnostic, privacy-preserving architecture for diverse organizational contexts.

\section*{Acknowledgements}{Funded by the European Union (TEADAL, 101070186). Views and opinions expressed are, however, those of the author(s) only and do not necessarily reflect those of the European Union. Neither the European Union nor the granting authority can be held responsible for them.}

%\bibliographystyle{IEEEtran}
%\bibliography{refs}

\begin{thebibliography}{10}
\providecommand{\url}[1]{#1}
\csname url@samestyle\endcsname
\providecommand{\newblock}{\relax}
\providecommand{\bibinfo}[2]{#2}
\providecommand{\BIBentrySTDinterwordspacing}{\spaceskip=0pt\relax}
\providecommand{\BIBentryALTinterwordstretchfactor}{4}
\providecommand{\BIBentryALTinterwordspacing}{\spaceskip=\fontdimen2\font plus
\BIBentryALTinterwordstretchfactor\fontdimen3\font minus
  \fontdimen4\font\relax}
\providecommand{\BIBforeignlanguage}[2]{{%
\expandafter\ifx\csname l@#1\endcsname\relax
\typeout{** WARNING: IEEEtran.bst: No hyphenation pattern has been}%
\typeout{** loaded for the language `#1'. Using the pattern for}%
\typeout{** the default language instead.}%
\else
\language=\csname l@#1\endcsname
\fi
#2}}
\providecommand{\BIBdecl}{\relax}
\BIBdecl

\bibitem{laux2024three}
J.~Laux, S.~Wachter, and B.~Mittelstadt, ``Three pathways for standardisation
  and ethical disclosure by default under the european union artificial
  intelligence act,'' \emph{Computer Law \& Security Review}, vol.~53, p.
  105957, 2024.

\bibitem{liu2020privacy}
C.~Liu, Q.~Zeng, L.~Cheng, H.~Duan, M.~Zhou, and J.~Cheng, ``Privacy-preserving
  behavioral correctness verification of cross-organizational workflow with
  task synchronization patterns,'' \emph{IEEE Transactions on Automation
  Science and Engineering}, vol.~18, no.~3, pp. 1037--1048, 2020.

\bibitem{liu2016towards}
C.~Liu, H.~Duan, Q.~Zeng, M.~Zhou, F.~Lu, and J.~Cheng, ``Towards comprehensive
  support for privacy preservation cross-organization business process
  mining,'' \emph{IEEE Transactions on Services Computing}, vol.~12, no.~4, pp.
  639--653, 2016.

\bibitem{yu2017survey}
X.~Yu, Z.~Yan, and A.~V. Vasilakos, ``A survey of verifiable computation,''
  \emph{Mobile Networks and Applications}, vol.~22, pp. 438--453, 2017.

\bibitem{smart2023computing}
N.~Smart, ``Computing on encrypted data,'' \emph{IEEE Security \& Privacy},
  vol.~21, no.~4, pp. 94--98, 2023.

\bibitem{bontekoe2023verifiable}
T.~Bontekoe, D.~Karastoyanova, and F.~Turkmen, ``Verifiable privacy-preserving
  computing,'' \emph{arXiv preprint arXiv:2309.08248}, 2023.

\bibitem{russinovich2024confidential}
M.~Russinovich, C.~Fournet, G.~Zaverucha, J.~Benaloh, B.~Murdoch, and M.~Costa,
  ``Confidential computing proofs: An alternative to cryptographic
  zero-knowledge,'' \emph{Queue}, vol.~22, no.~4, pp. 73--100, 2024.

\bibitem{zhang2021survey}
C.~Zhang, Y.~Xie, H.~Bai, B.~Yu, W.~Li, and Y.~Gao, ``A survey on federated
  learning,'' \emph{Knowledge-Based Systems}, vol. 216, p. 106775, 2021.

\bibitem{heiss_advancing_2022}
J.~Heiss, E.~Gr{\"u}newald, S.~Tai, N.~Haimerl, and S.~Schulte, ``Advancing
  blockchain-based federated learning through verifiable off-chain
  computations,'' in \emph{2022 IEEE International Conference on Blockchain
  (Blockchain)}.\hskip 1em plus 0.5em minus 0.4em\relax IEEE, 2022, pp.
  194--201.

\bibitem{lee_VDFL_2024}
C.~Lee, J.~Heiss, S.~Tai, and J.~W.-K. Hong, ``End-to-end verifiable
  decentralized federated learning,'' in \emph{2024 IEEE International
  Conference on Blockchain and Cryptocurrency (ICBC)}, 2024, pp. 434--442.

\bibitem{heiss_2021_trustworthyPreprocessing}
J.~Heiss, A.~Busse, and S.~Tai, ``Trustworthy pre-processing of sensor data in
  data on-chaining workflows for blockchain-based iot applications,'' in
  \emph{Service-Oriented Computing: 19th International Conference, ICSOC 2021,
  Virtual Event, November 22--25, 2021, Proceedings 19}.\hskip 1em plus 0.5em
  minus 0.4em\relax Springer, 2021, pp. 133--149.

\bibitem{lamb2021reproducible}
C.~Lamb and S.~Zacchiroli, ``Reproducible builds: Increasing the integrity of
  software supply chains,'' \emph{IEEE Software}, vol.~39, no.~2, pp. 62--70,
  2021.

\bibitem{munoz2023survey}
A.~Mu{\~n}oz, R.~Rios, R.~Rom{\'a}n, and J.~L{\'o}pez, ``A survey on the (in)
  security of trusted execution environments,'' \emph{Computers \& Security},
  vol. 129, p. 103180, 2023.

\bibitem{confidential2022common}
C.~C. Consortium \emph{et~al.}, ``Common terminology for confidential
  computing,'' \emph{Online],(December 2022).(Available from:
  https://confidentialcomputing.
  io/wp-content/uploads/sites/10/2023/03/Common-Terminology-for-Confidential-Computing.
  pdf)}, 2022.

\bibitem{brossard2023private}
M.~Brossard, G.~Bryant, B.~El~Gaabouri, X.~Fan, A.~Ferreira, E.~G. Evans,
  C.~Haster, E.~Johnson, D.~Miller, F.~Mo \emph{et~al.}, ``Private delegated
  computations using strong isolation,'' \emph{IEEE Transactions on Emerging
  Topics in Computing}, vol.~12, no.~1, pp. 386--398, 2023.

\bibitem{ben2019scalable}
E.~Ben-Sasson, I.~Bentov, Y.~Horesh, and M.~Riabzev, ``Scalable zero knowledge
  with no trusted setup,'' in \emph{Advances in Cryptology--CRYPTO 2019: 39th
  Annual International Cryptology Conference, Santa Barbara, CA, USA, August
  18--22, 2019, Proceedings, Part III 39}.\hskip 1em plus 0.5em minus
  0.4em\relax Springer, 2019, pp. 701--732.

\bibitem{bruestle2023risc}
J.~Bruestle and P.~Gafni, ``Risc zero zkvm: scalable, transparent arguments of
  risc-v integrity,'' \emph{Draft. July}, vol.~29, 2023.

\bibitem{wood2014ethereum}
G.~Wood, ``Ethereum: A secure decentralised generalised transaction ledger,''
  \emph{Ethereum project yellow paper}, vol. 151, no. 2014, pp. 1--32, 2014.

\bibitem{tara2019evolution}
A.~Tara, K.~Ivkushkin, A.~Butean, and H.~Turesson, ``The evolution of
  blockchain virtual machine architecture towards an enterprise usage
  perspective,'' in \emph{Software Engineering Methods in Intelligent
  Algorithms: Proceedings of 8th Computer Science On-line Conference 2019, Vol.
  1 8}.\hskip 1em plus 0.5em minus 0.4em\relax Springer, 2019, pp. 370--379.

\bibitem{IPFS_Benet_2014}
J.~Benet, ``{IPFS} - content addressed, versioned, {P2P} file system,''
  \emph{CoRR}, 2014.

\bibitem{eberhardt2018off}
J.~Eberhardt and J.~Heiss, ``Off-chaining models and approaches to off-chain
  computations,'' in \emph{Proceedings of the 2nd Workshop on Scalable and
  Resilient Infrastructures for Distributed Ledgers}, 2018, pp. 7--12.

\bibitem{xia2023poisoning}
G.~Xia, J.~Chen, C.~Yu, and J.~Ma, ``Poisoning attacks in federated learning: A
  survey,'' \emph{IEEE Access}, vol.~11, pp. 10\,708--10\,722, 2023.

\bibitem{nilsson2018performance}
A.~Nilsson, S.~Smith, G.~Ulm, E.~Gustavsson, and M.~Jirstrand, ``A performance
  evaluation of federated learning algorithms,'' in \emph{Proceedings of the
  second workshop on distributed infrastructures for deep learning}, 2018, pp.
  1--8.

\bibitem{moreau2011open}
L.~Moreau, B.~Clifford, J.~Freire, J.~Futrelle, Y.~Gil, P.~Groth,
  N.~Kwasnikowska, S.~Miles, P.~Missier, J.~Myers \emph{et~al.}, ``The open
  provenance model core specification (v1. 1),'' \emph{Future generation
  computer systems}, vol.~27, no.~6, pp. 743--756, 2011.

\bibitem{gehani2012spade}
A.~Gehani and D.~Tariq, ``Spade: Support for provenance auditing in distributed
  environments,'' in \emph{ACM/IFIP/USENIX International Conference on
  Distributed Systems Platforms and Open Distributed Processing}.\hskip 1em
  plus 0.5em minus 0.4em\relax Springer, 2012, pp. 101--120.

\bibitem{baracaldo2017securing}
N.~Baracaldo, L.~A.~D. Bathen, R.~O. Ozugha, R.~Engel, S.~Tata, and H.~Ludwig,
  ``Securing data provenance in internet of things (iot) systems,'' in
  \emph{Service-Oriented Computing--ICSOC 2016 Workshops: ASOCA, ISyCC, BSCI,
  and Satellite Events, Banff, AB, Canada, October 10--13, 2016, Revised
  Selected Papers 14}.\hskip 1em plus 0.5em minus 0.4em\relax Springer, 2017,
  pp. 92--98.

\bibitem{malik2018productchain}
S.~Malik, S.~S. Kanhere, and R.~Jurdak, ``Productchain: Scalable blockchain
  framework to support provenance in supply chains,'' in \emph{2018 IEEE 17th
  International Symposium on Network Computing and Applications (NCA)}.\hskip
  1em plus 0.5em minus 0.4em\relax IEEE, 2018, pp. 1--10.

\bibitem{mertens2024smart}
D.~Mertens, J.~Kim, J.~Xu, E.~Kim, and C.~Lee, ``Smart flow: a
  provenance-supported smart contract workflow architecture,'' \emph{Cluster
  Computing}, pp. 1--15, 2024.

\bibitem{ardagna2020blockchain}
C.~A. Ardagna, M.~Anisetti, B.~Carminati, E.~Damiani, E.~Ferrari, and
  C.~Rondanini, ``A blockchain-based trustworthy certification process for
  composite services,'' in \emph{2020 IEEE International Conference on Services
  Computing (SCC)}.\hskip 1em plus 0.5em minus 0.4em\relax IEEE, 2020, pp.
  422--429.

\bibitem{corradini2021chorchain}
F.~Corradini, A.~Marcelletti, A.~Morichetta, A.~Polini, B.~Re, F.~Tiezzi
  \emph{et~al.}, ``Chorchain: A model-driven framework for choreography-based
  systems using blockchain.'' in \emph{ITBPM@ BPM}, 2021, pp. 26--32.

\bibitem{kalka2024comprehensive}
M.~Kalka and M.~Kirejczyk, ``A comprehensive review of tlsnotary protocol,''
  \emph{arXiv preprint arXiv:2409.17670}, 2024.

\bibitem{delignat2023should}
A.~Delignat-Lavaud, C.~Fournet, K.~Vaswani, S.~Clebsch, M.~Riechert, M.~Costa,
  and M.~Russinovich, ``Why should i trust your code? confidential computing
  enables users to authenticate code running in tees, but users also need
  evidence this code is trustworthy.'' \emph{Queue}, vol.~21, no.~4, pp.
  94--122, 2023.

\bibitem{howard2023confidential}
H.~Howard, F.~Alder, E.~Ashton, A.~Chamayou, S.~Clebsch, M.~Costa,
  A.~Delignat-Lavaud, C.~Fournet, A.~Jeffery, M.~Kerner \emph{et~al.},
  ``Confidential consortium framework: Secure multiparty applications with
  confidentiality, integrity, and high availability,'' \emph{arXiv preprint
  arXiv:2310.11559}, 2023.

\bibitem{toldi2023blockchain}
B.~{\'A}. Toldi and I.~Kocsis, ``Blockchain-based, confidentiality-preserving
  orchestration of collaborative workflows,'' \emph{arXiv preprint
  arXiv:2303.10500}, 2023.

\bibitem{ramezan2024zk}
G.~Ramezan and E.~Meamari, ``zk-iot: Securing the internet of things with
  zero-knowledge proofs on blockchain platforms,'' \emph{arXiv preprint
  arXiv:2402.08322}, 2024.

\bibitem{heiss2023verifiable}
J.~Heiss, T.~Oegel, M.~Shakeri, and S.~Tai, ``Verifiable carbon accounting in
  supply chains,'' \emph{IEEE Transactions on Services Computing}, 2023.

\bibitem{zhou2011secure}
W.~Zhou, Q.~Fei, A.~Narayan, A.~Haeberlen, B.~T. Loo, and M.~Sherr, ``Secure
  network provenance,'' in \emph{Proceedings of the twenty-third ACM symposium
  on operating systems principles}, 2011, pp. 295--310.

\bibitem{zhou2022audiwflow}
X.~Zhou, A.~Nehme, V.~Jesus, Y.~Wang, M.~Josephs, K.~Mahbub, and A.~Abdallah,
  ``Audiwflow: Confidential, collusion-resistant auditing of distributed
  workflows,'' \emph{Blockchain: Research and Applications}, vol.~3, no.~3, p.
  100073, 2022.

\bibitem{taha2015trusted}
M.~M.~B. Taha, S.~Chaisiri, and R.~K. Ko, ``Trusted tamper-evident data
  provenance,'' in \emph{2015 IEEE Trustcom/bigdatase/ispa}, vol.~1.\hskip 1em
  plus 0.5em minus 0.4em\relax IEEE, 2015, pp. 646--653.

\end{thebibliography}

\end{document}